\documentclass{article}
 \usepackage[dvipdf]{epsfig}
  \usepackage{color}
\usepackage{graphicx}
\usepackage{color}
\def\be{\begin{equation}}
\def\ee#1{\label{#1}\end{equation}}
 \def\no{\nonumber}
\def\lb{\label}
\def\bx{\mathbf{x}}

\def\bv{\mathbf{v}}
\def\bp{\mathbf{p}}
\def\bk{\mathbf{k}}
\def\k{\textsf{k} }
\newcommand{\ben}{\begin{eqnarray}}
\newcommand{\een}{\end{eqnarray}}
\begin{document}

\title{\bf Analysis of instability of systems composed by  dark and baryonic matter}
\author{Gilberto  M. Kremer\footnote{kremer@fisica.ufpr.br} and Ra\'{\i}la Andr\'e\footnote{ra08@fisica.ufpr.br} \\
Departamento de F\'{\i}sica, Universidade Federal do Paran\'a, 81531-980 Curitiba, Brazil }
 \date{}
 \maketitle
\begin{abstract}
In this work  the dynamics of self-gravitating systems composed by dark and baryonic matter is analyzed. Searching for a description of this dynamics, a system of collisionless Boltzmann equations for the two constituents and the Poisson equation for the gravitational field are employed. Through the solution of these equations the collapse criterion is determined from  a dispersion relation. The collapse occurs in an unstable region where the solutions grow exponentially with time. It is shown that the unstable region  becomes larger if the dispersion velocity of dark matter becomes larger than the one of the baryonic matter.  The results obtained are also compared with the case where only the dark matter is present. The model of the present work  has a higher limit of instability and therefore, exhibited an advantage in the structure formation.
\end{abstract}




 \section{Introduction}

The problem of structure formation has been investigated for a long time, since the pioneer work of Jeans \cite{b1}. There are many models which explain the conditions for the beginning of the structure formation and when they occurred. The Jeans model describes the gravitational instability of a self-gravitating gas cloud and search for the conditions that small perturbations of a gas cloud grow exponentially, leading to the collapse of the cloud (see e.g. \cite{b2,b3,b4,b5,b6,b7,b8,b9,b10}). The simplest way to understand these criteria is to think in force balance arguments. The collapse occurs whenever the inwards directed gravitational force is larger than the outwards directed internal pressure of the gas. The parameters which are used to quantify this instability are the Jeans length and Jeans mass that will be discuss throughout this paper.

In this work the dark matter \cite{b11} is present together with baryonic matter. The accepted model of this unknown component is the model of cold dark matter. Such model postulates the dark matter as weakly interacting massive particles and whose typical velocity is much lower than that of the light. As a consequence, the cold dark matter is an auxiliary matter in a process of structure formation. In this scenario, the cold dark matter leads to a hierarchical formation process where small structures are formed first and massive structures later. Models with hot dark matter, on the other hand, show difficulties to explain the galaxies formation and other structures on a small scale. The higher velocities do not allow the agglomeration in the necessary scale. Thus, the model which will be analyzed in this work is composed by a mixture of two constituents, namely, a cold dark matter and a baryonic matter subjected to a gravitational field.

For the search of the collapse criteria for our model, we invoke the system of Boltzmann equations for the constituents and the Poisson equation for the gravitational field. This set of equations leads to a dispersion relation which implies in a  collapse criterion for infinity homogeneous fluid and stellar systems.

The aim of this work is: to obtain the Jeans length and the Jeans mass, to compare  the results with the case when only one constituent is present  and to discuss  the dispersion velocities ratio between baryonic and dark matter.

Some works with dark and baryonic matter were analyzed in the references \cite{b3,b4,b6,b10} by taking into account the macroscopic balance equations, but to the best of our knowledge the use of the system of Boltzmann equations  is new.

Another point of view without the inclusion of a dark matter constituent is to analyze the Jeans instability by using a $f(R)$- theory, which was the subject of investigation of the works \cite{cap,CL,RG}.

This paper is organized as follows. In section 2, we present the set of equations, the dispersion relation and we explore  the collapse limits  through the Jeans mass and Jeans length for a  system of collisionless baryonic and dark matter.   The conclusions of the work are stated in section 3.

\section{Systems composed of  collisionless dark and baryonic matter}

We start with the three equations that describe  systems composed by  dark and baryonic matter subjected to a gravitational field $\Phi$. Here we use the indices $d$ and $b$ for the  dark and baryonic matter, respectively. The evolution equations of the distributions functions of baryonic $f_b\equiv f(\bx,\bv_b,t)$ and dark matter $f_d\equiv f(\bx,\bv_d,t)$ are the collisionless Boltzmann equations (see Appendix)
\ben\lb{1a}
\frac{\partial f_b}{\partial t}+\bv_b\cdot\frac{\partial f_b}{\partial \bx} -\nabla\Phi\cdot\frac{\partial f_b}{\partial \bv_b}=0,\\\lb{1b}
\frac{\partial f_d}{\partial t}+\bv_d\cdot\frac{\partial f_d}{\partial \bx} -\nabla\Phi\cdot\frac{\partial f_d}{\partial \bv_d}=0.
\een
 The gravitational field must  fulfill the Poisson equation:
\ben\lb{2}
\nabla^2\Phi=4\pi G \left(\int f_b d\bv_b+\int f_d d\bv_d\right)
=4\pi G(\rho_b+\rho_d),
\een
where $\rho_b$ and $\rho_d$ are the mass densities of the baryonic and dark matter, respectively.

We consider that the equilibrium of the self-gravitating  system is described by  homogeneous and time-independent distribution functions $f_b^{0}(\bv_b)$, $f_d^{0}(\bv_d)$ and a potential which is only a function of the space coordinates, i.e.,  $\Phi_{0}(\bx)$.   The equilibrium state is then subjected to small perturbations represented by plane waves of frequency $\omega$ and wavenumber vector $\bk$, namely
\ben\lb{3a}
f(\bx,\bv_b,t)=f_b^{0}(\bv_b)+\overline{f}_b^1\exp[i(\bk\cdot\bx-\omega t)],,
\\\lb{3b}
f(\bx,\bv_d,t)=f_d^{0}(\bv_d)+\overline{f}_d^1\exp[i(\bk\cdot\bx-\omega t)],,
\\\lb{3c}
\Phi(\bx,t)=\Phi_{0}(\bx)+\overline{\Phi}_1\exp[i(\bk\cdot\bx-\omega t)],
\een
 where the overbarred quantities denote small amplitudes.

 The equilibrium for a homogeneous system is achieved by Jeans "swindle" that allows us to make $\Phi_0=0$  without loss of consistency. Hence, if we insert (\ref{3a}) -- (\ref{3c}) into the system of equations of (\ref{1a}) -- (\ref{2}) and  linearize the resulting equations, we get
\ben\lb{4a}
-i\omega \overline{f}_b^1+\bv_b\cdot \left(i\bk \overline{f}_b^1\right)-\left(i\bk\overline{\Phi}_1\right)\cdot \frac{\partial f_b^0}{\partial \bv_b}=0,
\\\lb{4b}
-i\omega \overline{f}_d^1+\bv_d\cdot \left(i\bk \overline{f}_d^1\right)-\left(i\bk\overline{\Phi}_1\right)\cdot \frac{\partial f_d^0}{\partial \bv_d}=0,
\\\lb{4c}
-\k^2\overline{\Phi}_1=4\pi G\left(\int \overline{f}_b^1 d\bv_b+\int \overline{f}_d^1 d\bv_d\right).
\een
Here  the modulus of the wavenumber vector was denoted by $\k=\vert\bf k\vert$.

By eliminating the overbarred quantities from the system of equations (\ref{4a}) -- (\ref{4c}) it follows the dispersion relation
\ben\lb{5}
\k^2+4\pi G\,\bk\cdot\left(\int\frac{\partial f_b^0}{\partial \bv_b}\frac{d\bv_b}{\bv_b\cdot\bk-\omega}
+\int\frac{\partial f_d^0}{\partial \bv_d}\frac{d\bv_d}{\bv_d\cdot\bk-\omega}\right)=0.
\een

From now on we assume that the distribution functions $f_b^0$ and $f_d^0$ are the Maxwellians:
\ben\lb{6}
f_b^0(\bv_b)=\frac{\rho_b^0}{ (2\pi\sigma_b^2)^\frac32 }e^{-{v_b^2}/{2\sigma_b^2}},\qquad
\qquad
f_d^0(\bv_d)=\frac{\rho_d^0}{ (2\pi\sigma_d^2)^\frac32 }e^{-{v_d^2}/{2\sigma_d^2}},
\een
where $\rho_b^0$,  $\rho_d^0$ are constant mass densities of baryons and dark matter, respectively, and $\sigma_b$ and $\sigma_d$    their  dispersion  velocities.

Without loss of generality we can choose $\bk=(\k,0,0)$ so that the dispersion relation  (\ref{5}) together with (\ref{6}) can be integrated with respect to the velocity components $v_y$ and $v_z$, yielding
\ben\lb{7a}
\k^2=4\pi G\frac2{\sqrt{\pi}}\left[\frac{\rho_b^0}{\sigma_b^2}\int_0^\infty \frac{x^2 e^{-x^2}}{x^2-w^2/(2\sigma_b^2\k^2)}+\frac{\rho_d^0}{\sigma_d^2}\int_0^\infty \frac{y^2 e^{-y^2}}{y^2-w^2/(2\sigma_d^2\k^2)}\right].
\een
Above, we have introduced new integration variables
\ben\lb{7b}
x=\frac{v_{bx}}{\sqrt{2}\,\sigma_b},\qquad
y=\frac{v_{dx}}{\sqrt{2}\,\sigma_d}.
\een

Unstable solutions are such that $\Re( {\omega})=0$ and $\omega_I=\Im(\omega)>0$, since in this case the solutions grow exponentially with time. When $\omega=i\omega_I$ the integrals on the right-hand side of (\ref{7a}) can be evaluated by using the following relationship (see \cite{GR} eq. 3.466)
\ben\lb{8a}
\int_0^\infty \frac{x^2 e^{-\mu^2x^2}}{x^2+\beta^2}dx=\frac{\sqrt\pi}{2\mu}-\frac{\pi\beta}{2}e^{\beta^2}\,{\rm
erfc}(\beta\mu),
\een
where $\rm erfc$ is the complementary error function.
By taking into account (\ref{8a}) the dispersion relation (\ref{7a}) with $\omega=i\omega_I$ reduces to
\ben\no
\k_\ast^2&=&1-\sqrt{\frac\pi2}\frac{\omega_\ast}{\k_\ast}\exp\left({{\frac{\omega_\ast^2}{2{\k}_\ast^2}}}
\right){\rm erfc}\bigg(\frac{\omega_\ast}{\sqrt{2}\k_\ast}\bigg)
\\\lb{8b}
&+&\frac{\rho_b^0\sigma_d^2}{\rho_d^0\sigma_b^2}\bigg[1-\sqrt{\frac\pi2}\frac{\sigma_d}{\sigma_b}\frac{\omega_\ast}{\k_\ast}
\exp\left({\frac{\sigma_d^2}{\sigma_b^2}\frac{\omega_\ast^2}{2{{\k}}_\ast^2}}\right){\rm erfc}\bigg(\frac{\sigma_d}{\sigma_b}\frac{\omega_\ast}{\sqrt2\k_\ast}\bigg)\bigg].\qquad
\een
In the above equation we have introduced the dimensionless wavenumber and the dimensionless frequency defined by
\ben\lb{8c}
\k_\ast=\frac{\sigma_d\k}{\sqrt{4\pi G\rho_d^0}},\qquad \omega_\ast=\frac{\omega_I}{\sqrt{4\pi G\rho_d^0}}.
\een
Note that we have used the mass density $\rho_d$ and the dispersion velocity $\sigma_d$ of the dark matter in order to  construct the  dimensionless wavenumber and frequency, since the dark matter plays an important  role in the cosmic structure formation.

From (\ref{8b}) we may determine the instability regions in the plane $(\omega_\ast,\k_\ast)$, where disturbances grow with time. For that end we have to specify: (a) the mass densities ratio $\rho_d^0/\rho_b^0$ and (b) the dispersion velocities ratio $\sigma_d/\sigma_b$. The mass densities ratio can be taken as the ratio of the densities parameters $\Omega_d/\Omega_b$ today which is approximately 5.5 (see the section 22 of \cite{PP}), i.e., $\rho_d^0/\rho_b^0\approx5.5$. This ratio  has not modified  too much during the evolution of the Universe. For the dispersion velocities ratio we take the values given in the work  \cite{Sim} on the Milky-Way-like galaxy simulations
including both dark matter and baryons, namely,  $\sigma_d/\sigma_b=170/93\approx1.83$.

By considering $\omega_\ast=0$ in the dispersion relation (\ref{8b}),  we obtain that the  dimensionless  wavenumber is equal to $\k_* = 1.2694$. This value can be interpreted as the ratio of the Jeans wavenumbers one  referring to the dark matter-baryons system $\k_J^{db}$ and the other only to dark matter $\k_J^d={\sqrt{4\pi G\rho_d^0}}/\sigma_d$. Hence, the ratio of the Jeans wavelengths $\lambda_J^{db}=2\pi/\k_J^{db}$ and $\lambda_J^d=2\pi/\k_J^d$ is $\lambda_J^{db}/\lambda_J^{d}=0.7877$, showing that the Jeans wavelength of the system dark matter-baryons is smaller than the one with only the dark matter. This reflects also in the value of the Jeans mass, which is defined as the mass contained in a sphere of diameter equal to the Jeans wavelength. The ratio of the Jeans masses of the systems dark matter-baryons $M_J^{db}$ and only dark matter $M_J^d$ is
\ben\lb{9}
\frac{M_J^{db}}{M_J^{d}}=\frac{\rho^0_d+\rho^0_b}{\rho^0_d}\left(\frac{\lambda_J^{db}}{\lambda_J^{d}}\right)^3
=\frac{\rho^0_d+\rho^0_b}{\rho^0_d}\sqrt{\left(\frac{\rho^0_d\sigma_b^2}{\rho^0_d\sigma_b^2+\rho^0_b\sigma_d^2}
\right)^3}=0.5791.
\een
This shows that in the model where baryonic and dark matter  are present, the structures began to form earlier, at the time that the dark component dominated, which reinforces the fact that a smaller Jeans mass was required to initiate the collapse.

\begin{figure}[ht]\
\includegraphics[width=0.45\textwidth]{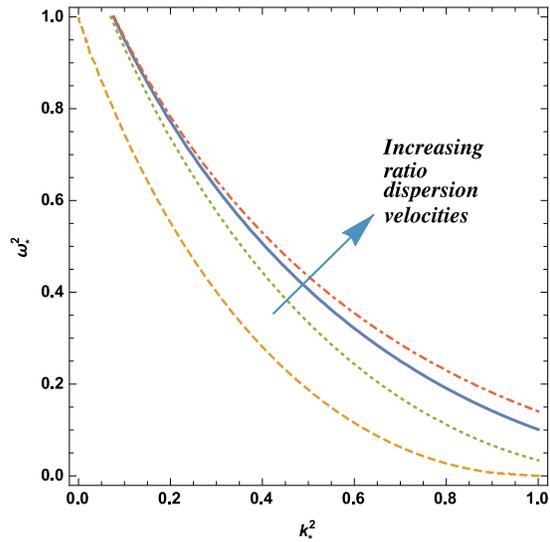}
\caption{Dimensionless wavenumber versus dimensionless frequency.  Models with baryonic and dark matter with: (a)   $\sigma_d/\sigma_b=1.83$ thick line; (b)  $\sigma_d/\sigma_b=1.20$ dotted line; (c)  $\sigma_d/\sigma_b=2.2$ dot-dashed line. Model
 with only dark matter dashed line. }
\lb{Fig3}
\end{figure}

We may ask about the role of the dispersion velocity ratio on the structure formation. This can be seen from Fig. \ref{Fig3} where three different values for $\sigma_d/\sigma_b$ were considered. The regions below the lines refer to the positive values of the contour plot of (\ref{8b}), while the regions above correspond to the  negative values. The arrow in this figure indicates the direction of the increase of the dispersion velocities ratio. Here we may infer that by increasing the dispersion velocity ratio the unstable region also increases.  As a consequence the ratios of the Jeans masses become: (a) ${M_J^{db}}/{M_J^{d}}=0.8338$ for $\sigma_d/\sigma_b=1.20$   and (b) ${M_J^{db}}/{M_J^{d}}=0.4585$ for  $\sigma_d/\sigma_b=2.2$. Hence, the increase of the dispersion velocity ratio implies that a smaller  mass is needed to begin the collapse.

\section{Conclusions}

In this work, we analyzed the dynamics and the collapse of collisionless self-gravitating system composed by dark and baryonic matter. This system is described by two  Boltzmann equations, one for each constituent, and the Poisson equation for the gravitational field. The dispersion relation for this model was found by linearizing the equations about the equilibrium of the self-gravitating system, which  is described by homogeneous and
time-independent Maxwellian distribution functions  and a potential which obeys Jeans "swindle".  The dispersion relation allows to obtain the collapse criterion of interstellar gas clouds composed by the mentioned constituents, resulting in the formation of structures.

The Jeans mass always has lower values in the models where the baryonic and dark matter are present. This phenomena occurs because the dark matter agglomerates more easily and the baryonic matter passes to aggregate due to the attraction of the gravitational field generated from this initial agglomerate. This result demonstrates that the structures began to form earlier when the dark matter dominated. We also conclude that when the dispersion velocity  of the baryonic particles is relatively smaller than the one of the dark matter, they  easily aggregate, since they hardly overcome the escape velocity of a given gravitational field. This behavior can be observed through the Fig. \ref{Fig3}, when $\sigma_{d} / \sigma_{b}$ increases, the unstable region -- where the solutions grow exponentially -- increases also.

Although the model described here is a function of two ratios -- namely, of the densities and of the dispersion velocities of baryonic and dark matter -- and the values of these ratios must be specified in order to analyze Jeans instability, we think that it is appropriate to analyze collisionless self-interacting systems of dark and baryonic matter. Here a kinetic model with two Boltzmann equations one for each constituent coupled with the Poisson equation was considered. It is the kinetic counterpart of the phenomenological model which considers the balance equations  of mass and momentum for each constituent coupled with the Poisson equation (see e.g. \cite{b3,b6}).

To sum up, this model with baryonic and dark matter proved to have a higher limit of instability and  therefore, exhibited an advantage in the structure formation.

\section*{Acknowledgments}

The authors acknowledge the Conselho Nacional de Desenvolvimento Cient\'{\i}fico e Tecnol\'ogico (CNPq), Brazil, for financial support.

\appendix

\section{The Boltzmann equation}
A state of a non-relativistic gas mixture of $r$ constituents is characterized by the set of one-particle distribution functions $f_a\equiv f(\bx,\bv_a,t), (a=1,\dots ,r)$ in the phase  space spanned by $(\bx,\bv_a)$, such that $f(\bx,\bv_a,t)d{\bf x} d{\bf v}_a$ gives the total mass of particles of constituent $a$ in the volume element $d{\bf x}$ about the position $\bx$ and with velocity in the range $d{\bf v}_a$ about $\bv_a$. The space-time evolution of the one particle distribution function $f_a$  in the phase space is governed by the Boltzmann equation, which in the presence of a gravitational potential $\Phi$ and in the absence of collisions between the particles, reads
(see e.g. \cite{GMK})
\ben\lb{A1}
\frac{\partial f_a}{\partial t}+\bv_a\cdot\frac{\partial f_a}{\partial \bx} -\nabla\Phi\cdot\frac{\partial f_a}{\partial \bv_a}=0,\qquad a=1,\dots r,\qquad
\een
where the force which acts on a particle of constituent $a$ is only of gravitational nature ${\bf F}=-\nabla\Phi$.

The above equation follows also from a relativistic Boltzmann equation in the presence of gravitational fields, where a one-particle distribution function $f_a\equiv f(\bx,\bp_a,t)$ of constituent $a$ in the phase space spanned by $(\bx,\bp_a)$ satisfies the Boltzmann equation (see e.g. \cite{CK})
\ben\lb{A2}
p_{a}^{\mu}\frac{\partial f_{a}}{\partial x^{\mu}}-\Gamma_{\mu\nu}^{i}p_{a}^{\mu}p_{a}^{\nu}\frac{\partial f_{a}}{\partial p_{a}^{i}}=0,\qquad a=1,\dots r.
\een
Here the mass-shell condition $p_a^\mu p_{a\mu}=m_a^2c^2$ -- where $m_a$ is the particle rest mass of constituent $a$ -- was taken into account. In the non-relativistic  Newtonian limiting case $p_a^0 \rightarrow m_ac$, $\bp_a\rightarrow m_a\bv_a$ and the Christoffel symbol $\Gamma_{00}^i\rightarrow\nabla^i\Phi/c^2$ so that (\ref{A1}) follows.

The equilibrium  distribution function is a Maxwellian distribution of the velocities
\ben\lb{A3}
f_a^0(\bx,\bv_a,t)=\frac{\rho_a}{ (2\pi\sigma_a^2)^\frac32 }e^{-{v_a^2}/{2\sigma_a^2}},
\een
where $\sigma_a=\sqrt{kT_a/ m_a}$ -- with $k$ denoting the Boltzmann constant and $T_a$ the temperature of the constituent $a$ -- is the dispersion velocity and $\rho_a$  the mass density of constituent $a$, which is defined in terms of the one-particle distribution function by
\ben\lb{A4}
\rho_a=\int f_a d{\bf v}_a.
\een


\begin{thebibliography}{0}
\bibitem{b1} J. H. Jeans, \emph{ Phil. Trans.  Royal Soc. London}  {\bf199}  (1902) 1.
\bibitem{b2} S. Weinberg,  \emph{Gravitation and Cosmology, principles and applications of the general theory of relativity} ( John Wiley \& Sons, New York, 1972).
\bibitem{b3}  J. A. Peacock, \emph{Cosmological Physics} (Cambridge University Press, Cambridge, 1999).
\bibitem{b4} P. Coles and F. Lucchin,  \emph{The origin and evolution of cosmic structure}
2nd. edn. (John Wiley, Chichester,  2002).
\bibitem{b5} J.  Binney and S. Tremaine,  \emph{Galatic Dynamics} 2nd. edn. (Princeton University Press, Princeton, 2008).
\bibitem{b6} M. L. Longair, \emph{Galaxy Formation} 2nd. edn, (Springer-Verlag, Berlin, 2008).
\bibitem{b7}  W. B. Bonnor, \emph{MNRAS} {\bf117}  (1957) 104.
\bibitem{b8} C. Low and D. Lynden-Bell,  \emph{MNRAS} {\bf176}  (1976) 367.
\bibitem{b9} J. M. Owen and J. V. Villumsen,  \emph{ApJ} {\bf481}   (1997) 1.
\bibitem{b10} D. Tsiklauri,  \emph{ApJ} {\bf507}   (1998) 226.
\bibitem{b11} G. Bertone D. Hooper and  J. Silk,  \emph{Phys. Rep.} {\bf405}  (2005) 279.
\bibitem{cap} S. Capozziello,   M. De Laurentis, I. De Martino, M. Formisano and S.D. Odintsov,  \emph{Phys. Rev. D} {\bf 85}  (2012) 044022.
\bibitem{CL} S. Capozziello and   M. De Laurentis, \emph{Ann. Phys.} {\bf524} (2012) 545.
\bibitem{RG} R. Andr\'e and G. M. Kremer, \emph{astro-ph1411.6096v1}.
\bibitem{GR} I. S. Gradshteyn and I. M.  Ryzhiz, \emph{Tables of Integrals, Series and Products} 7th edn.
 (Academic Press, Burlington,  2007).
\bibitem{PP} K. A. Olive et al. (Particle Data Group), \emph{Chin. Phys. C} {\bf38} (2014) 090001.
\bibitem{Sim} F.-S. Ling, E. Nezri, E. Athanassoulab and R. Teyssie, \emph{ JCAP} 02 (2010) 012.
\bibitem{GMK} G. M. Kremer,  \emph{An Introduction to the Boltzmann Equation and Transport Processes in Gases} (Springer-Verlag, Berlin, 2010).
\bibitem{CK} C. Cercignani and G. M. Kremer, \emph{The Relativistic Boltzmann Equation: Theory and Applications} (Birkh\"auser, Basel, 2002).

\end{thebibliography}
\end{document}